\documentclass[12pt,twoside]{article}
\usepackage[super,sort,comma]{natbib}
\usepackage{fancyhdr, float} %Gives headers and footers defined below
\newcommand{\captionv}[3]{\begin{center}\parbox{#1cm}{\caption[#2]{{\sf #3}}}
        \end{center}}
\usepackage[section]{placeins}
\usepackage{graphicx}

\makeatletter \renewcommand\@biblabel[1]{$^{#1}$} \makeatother
\usepackage[mathlines]{lineno}

\usepackage{hyperref}
\hypersetup{ colorlinks,
    citecolor=blue,
    filecolor=blue,
    linkcolor=blue,
    urlcolor=blue
}

\usepackage{xcolor}
\definecolor{gray}{rgb}{0.6,0.6,0.6}
\definecolor{red}{rgb}{0.85,0,0}
\definecolor{green}{rgb}{0,0.85,0}
\definecolor{blue}{rgb}{0,0,0.85}
\definecolor{beige}{rgb}{0.92,0.87,0.78}

\usepackage[all]{hypcap} %causes the link to figures to go to figure, not caption
\begin{document}

%\title{Effect of uncertainties in the dose rate curves in molecular radiotherapy on tumor control probability}
\begin{center}
    \sf {\Large {\bfseries Analysis of the sensitivity of tumor control probability in molecular radiotherapy to uncertainties in the dose rate curves} \\
    \vspace*{10mm}
    Pedro Otero-Casal$^{1, 2, +}$, Aldán Baliño$^{1, 2, +}$, Sara Neira$^{1,++}$, Faustino Gómez$^2$, Juan Pardo-Montero$^{1, 3, *}$} \\
    $^1$Group of Medical Physics and Biomathematics, Instituto de Investigación Sanitaria de Santiago (IDIS), Santiago de Compostela, Spain; $^2$Department of Particle Physics, Universidade de Santiago de Compostela, Santiago de Compostela, Spain; $^3$Department of Medical Physics, Complexo Hospitalario Universitario de Santiago, Santiago de Compostela, Spain.
    \vspace{5mm}\\
    Version typeset \today\\
\end{center}
%\maketitle

\pagenumbering{roman}
\setcounter{page}{1}
\pagestyle{plain}

*{\bf Correspondence to:} Juan Pardo-Montero, Grupo de Física Médica e Biomatemáticas, Instituto de Investigación Sanitaria de Santiago (IDIS), Servizo de Radiofísica e Protección Radiolóxica, Hospital Clínico Universitario de Santiago, Trav. Choupana s/n, 15706, Santiago de Compostela (Spain); E-mail:  juan.pardo.montero@sergas.es \\

+ Equal contribution. \\
++ Now at Department of Digitalization, Reganosa Servicios, Mugardos, Spain. 

{\bf Short title: Dose rate uncertainties and TCP in molecular radiotherapy} 
\newpage

\begin{abstract}

%\noindent 
{\bf Background:} The response to a molecular radiotherapy (MRT) treatment depends on the dose and dose rate. The reconstruction of the dose rate curves presents uncertainties that can affect the quantification of the radiobiological effect. \\ 

{\bf Purpose:} To investigate the sensitivity of the effectiveness (TCP) of an MRT treatment to uncertainties of the dose rate curves that may appear when reconstructing those curves from limited data samples.\\

{\bf Methods:} We generated different dose rate curves from experimental data, imposing the constraint of equal dose for each of them. Then, we computed TCPs and looked for correlations between metrics measuring the differences between the dose rate curves and differences in TCP. Finally, according to these results, we estimated the range of tolerable uncertainties in the dose rate curves (those that lead to differences in TCP below a certain clinical threshold). The study was performed for different radiopharmaceuticals (with different decay/clearance rates), 131I-CLR1404, 90Y-DOTATOC and 64CuCl2, and different radiosensitive parameters that can affect the dose rate response ($\alpha/\beta$, sub-lethal repair rate).\\

{\bf Results:} The best correlation between differences in the dose rate curves and TCP was found for a metric that computes averaged linear differences between the curves (Spearman coefficient 0.72). With this metric, we quantified differences in  dose rate curves that would lead to differences in TCP of 0.02, a parameter denoted $m_{1,\: 0.02}$ that is a surrogate of the dependence of the TCP on the dose rate profile. The results showed that the sensitivity of the TCP to dose rate variations decreases (\emph{i.e.} larger values of $m_{1,\: 0.02}$) with increasing $\alpha/\beta$ and sub-lethal damage repair rate of the tumor cells, and increasing biological half-life of the dose rate curves. For example, values of $m_{1,\: 0.02}$ were around 8$\times$10$^{-3}$ (5$\times$10$^{-3}$), 15$\times$10$^{-3}$ (9$\times$10$^{-3}$), and 85$\times$10$^{-3}$ (28$\times$10$^{-3}$) for 64CuCl2, 90Y-DOTATOC, and 131I-CLR1404, respectively, with $\alpha/\beta$=10 Gy (3 Gy), and sub-lethal damage repair rate 0.69 h$^{-1}$ in all cases.\\

{\bf Conclusions:} The radiobiological effect of a MRT treatment on a tumor depends not only on the absorbed dose but also on the dose rate profile. Ideally, both magnitudes should be measured with accuracy in order to progress towards the optimization of individualized MRT treatments. Our study suggests that this would be more important for tumors with low $\alpha/\beta$ and moderately slow sub-lethal damage repair ($\sim$0.5 h$^{-1}$) treated with fast-decaying radiopharmaceuticals.

\end{abstract}

% The table of contents is for drafting and refereeing purposes only.
%\tableofcontents

\newpage

\setlength{\baselineskip}{0.7cm} %double spacing		

\pagenumbering{arabic}
\setcounter{page}{1}
\pagestyle{fancy}

\section[Introduction]{Introduction}
\label{section_intro}

Molecular radiotherapy (MRT; also commonly known as targeted radionuclide therapy, TRT) employs different radiopharmaceuticals to treat cancer \citep{sgouros2020}. Radiopharmaceuticals are typically administered to the patient orally or intravenously, where they are metabolized and preferentially uptaken by the tumor/s. The decay of the radionuclides causes the irradiation of the tumor. Different radionuclides emitting short-range particles are employed in MRT to treat different types of cancer, among them $^{131}$I, $^{177}$Lu or $^{90}$Y \citep{silberstein2012, fendler2017, cwikla2010}.

The radiation dosimetry of patients undergoing MRT is far from the accuracy achieved in external radiotherapy due to the systemic nature of this therapy. Absorbed dose is associated with tumor control and toxicity in radiotherapy, and therefore an accurate determination of the absorbed dose is paramount to design optimal individualized MRT treatments based on the dose-effect relationship, as it is done in external radiotherapy \citep{davis2023, stokke2017}. Dosimetry in MRT is a very active field of research, aiming at improving the dosimetric accuracy and moving a step closer to dose-based individualized treatment planning\cite{ivashchenko2024, strigari2014}. Improved methods include pharmacokinetic modelling to improve the reconstruction of time-activity curves (TACs)\cite{jeremic2018, guiu2018, neira2021}, the use of Monte Carlo calculations to compute the absorbed dose \cite{kost2015,marcatili2013,besemer2018,neira2020b}, and proposals to experimentally monitor the spatio-temporal evolution of the activity in the patients through the use of systems of external detectors \cite{morganti2021,mancini2023}.

In general, emphasis is on the accurate calculations of the absorbed dose, not on the accurate reconstruction of the dose rate curves. Dose rate curves are related to time-activity curves, and the accurate reconstruction of the TACs would require multiple experimental measurements, which may not be clinically feasible. Some studies have shown that the reconstruction of the dose rate (TAC) presents more uncertainties than the reconstruction of the absorbed dose (integral of the TAC), because of the averaging of the integral (different TACs can lead to the same integrated activity) \cite{neira2020b, morganti2021}.

Nonetheless, it is well known that the dose rate plays an important role in the response to MRT\citep{dale1996, dale2005}. Sub-lethal damage repair during irradiation can decrease the effect of a radiation dose, $D$. This is typically accounted for by including a sub-lethal repair term through a protraction factor, $G$, that affects the quadratic term of the linear-quadratic (LQ) model \cite{orourke2009, steel1986, dale2005}:

\begin{equation}
\log	 SF = -\alpha D -\beta G D^2
\label{eq1}
\end{equation}

The protraction factor depends on the dose rate profile and damage/repair parameters of the cells and fulfils that 0$\leq$$G$$\leq$1, with $G\to1$ for instantaneous dose delivery (and Eq. (\ref{eq1}) approaching the LQ model) and $G\to0$ for very prolonged dose deliveries. This means that radiation-induced cellular death may depend not only on the total radiation dose administered, but also on the dose-rate pattern of the radiation delivery. This is known as the dose-rate effect, which has been deeply studied in the literature of radiobiological modelling \cite{dale1985,  hall1991, dale1999, dale2005}.

Recently, Galler \emph{et al.} have theoretically investigated the effect of dose rate on the tumor control achieved with multi-cycle $^{177}$Lu MRT \cite{galler2024}. They used the mechanistic lethal-potentially lethal model to simulate the dynamics of radiation-induced cell lesions and the Zaider-Minerbo model to compute the tumor control probability (TCP) including repopulation of tumor cells for different dose rate patterns and radiobiological parameters. They showed that different dose rate profiles result in significantly different TCPs, even if the total dose absorbed by the tumor remains constant. This dose rate effect is stronger when the repair of sub-lethal damage is slow (lower repair rate). 

In this work, we have investigated how sensitive the effectiveness of an MRT treatment is to uncertainties of the dose rate curves that may appear when reconstructing those curves from limited data samples. The study has been performed for different radiopharmaceuticals (with different decay/clearance rates) and different radiosensitive parameters that can affect the dose rate response ($\alpha/\beta$, sub-lethal repair rate). For this purpose, we generated different dose rate curves from experimental data, imposing the constraint of equal dose for each of them. Then, we computed TCPs and looked for correlations between metrics measuring the differences between the dose rate curves and differences in TCP. Finally, according to these results, we estimated the range of tolerable uncertainties in the dose rate curves (those that lead to differences in TCP below a certain clinical threshold) according to the clearance rate of the dose rate curve and the above-mentioned radiobiological parameters.

\section[Materials and Methods]{Materials and Methods}
\label{section_materials}

\subsection[]{Radiobiological modelling}
\label{section_2_1}

We relied on the LQ model with sub-lethal repair (Eq. (\ref{eq1})) to model response. Assuming that sub-lethal damage repair follows an exponential with repair rate $\mu$, the dose protraction factor for an arbitrary dose rate pattern $\dot{D}(t)$ can be calculated as:

\begin{equation}
G = 2 \int_0^T \frac{\dot{D}(t)}{D}{\rm d}t \int_0^t \frac{\dot{D}(t^\prime)}{D} e^{-\mu (t-t^\prime)} {\rm d}t^\prime	
\label{eq2}
\end{equation}

In the equation above $D$ is total dose ($D=\int_0^T \dot{D}(t)dt$) and $T$ is the delivery time. In MRT, the delivery time is formally $\infty$, and a value $T$ at which the dose rate has decayed enough to not introduce a bias in the computation of $G$ was used (different for each radiopharmaceutical under investigation, as discussed in section \ref{section_2_1}).

We used the Poisson-LQ approach to calculate tumor control probabilities \cite{webb1993}:
\begin{equation}
    \mathit{TCP}= \exp(-N\mathit{SF}),
    \label{eq3}
\end{equation}
where $N$ is the number of clonogenic cells within the tumor at the irradiation time and $\mathit{SF}$ is the surviving fraction after a treatment, obtained from Equations~(\ref{eq1}) and (\ref{eq2}). It is well known that this expression leads to dose-response curves that are too steep unless sensitivity averaging is included \cite{webb1993}. We have included averaging in the parameters $\alpha$, $\beta$, $\mu$ and $N$: 

\begin{equation}
    \mathit{TCP}= \int_N \exp \left(-N \left(\int_{\alpha,\beta,\mu}  \mathit{SF} ~ f_{\alpha} ~ f_{\beta} ~ f_{\mu} ~ {\rm d}\alpha ~ {\rm d}\beta ~{\rm d}\mu \right) \right) f_N ~ {\rm d}N
    \label{eq4}
\end{equation}

In the above equation the functions $f$ represent the probability distribution functions of each parameter. We used normal distributions for each of them with \emph{variance}=$(0.2 ~ \mathit{mean})^2$.

\subsection[]{Experimental data and generation of dose rate curves for MRT}
\label{section_2_2}

We focused our study on three radiopharmaceuticals with very different biokinetics, 131I-CLR1404, 90Y-DOTATOC and 64CuCl2. For them, we obtained experimental biokinetic data in the tumor from Refs. \cite{besemer2019, jeremic2018, righi2018}.

\begin{figure}[t]
    \centering
    \includegraphics[width=10cm]{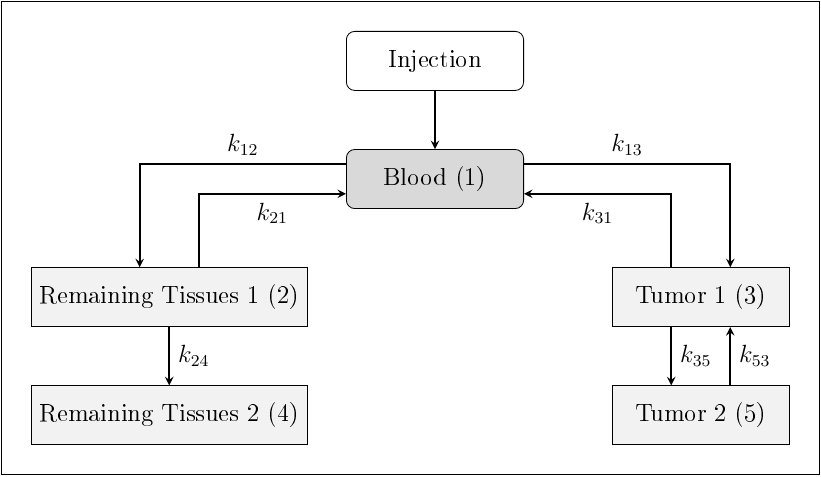}
    \captionv{14}{}{\label{fig_1} Graphical sketch of the biokinetic model used to generate different time activity curves in the tumor.}
\end{figure}

A simple pharmacokinetic model with 5 compartments was used to generate the TACs of the different radiopharmaceuticals from the experimental data points (Figure \ref{fig_1}). Mathematically, the system of equations describing the model is given by

\begin{equation}
\frac{dy_i}{dt} = \sum_j k_{ji}y_j - \sum_i k_{ij}y_i-\lambda y_i
\label{eq5}
\end{equation}
where $y_i$ is the time-activity curve of the compartment $i$, $k_{ij}$ is the coupling constant between compartments $i$ and $j$ and $\lambda$ is the decay rate of the radioisotope.

A reference $\mathit{TAC}_0$ was obtained by fitting all the experimental data points. The cumulated activity of this TAC, $\mathit{CA}_0$, was computed, and then different TACs ($\mathit{TAC}_i$, $i>0$) were computed by fitting the model to subsets of experimental points while imposing a constraint of the cumulated activity of the TAC to match the cumulated activity of the reference TAC ($\mathit{CA}_i=\mathit{CA}_0$). This was done independently for 131I-CLR1404, 90Y-DOTATOC and 64CuCl2.

Dose rate curves in the tumor were obtained simply by scaling the TACs as $\dot{D}_i(t)=a \mathit{TAC}_i$. Certainly, this approximation ignores contributions to the dose rate in the tumor coming from the decay of radionuclides located in organs/tissues, but that is not a limitation for the purpose of this work. The scaling factor, $a$, was calculated for each radiopharmaceutical under investigation and each combination of radiobiological parameters to yield a TCP of 50\% (Eq. (\ref{eq4})) for the reference curve (this calculation was performed by implementing a bisection algorithm with stopping condition $0.498<\mathit{TCP}_{\rm ref}<0.502$).

\subsection[]{Quantification of differences between dose rate curves and correlations with differences in TCP}

We investigated four different metrics to quantify differences between dose rate curves in order to investigate correlations between dose rate differences (metric) and radiobiological differences (TCP). These metrics, presented below, quantify the difference between a dose rate curve, $\dot{D}_i(t)$, and a reference curve $\dot{D}_0(t)$ by computing global (averaged) linear/quadratic differences and local maximum differences between the curves as: 

\begin{equation}
m_1^i = \overline{\left(	\frac{|\dot{D}_0(t) - \dot{D}_i(t)|}{\max(\dot{D}_0(t)) } \right)};\;\; m_2^i = \overline{\left(	\frac{\left(\dot{D}_0(t) - \dot{D}_i(t)\right)}{\left(\max(\dot{D}_0(t))\right)} \right)^2};\;\; m_3^i = \max \left(	\frac{|\dot{D}_0(t) - \dot{D}_i(t)|}{\max(\dot{D}_0(t)) } \right)
\end{equation}

%\;\;\; m_4^i = \max \left(	\frac{\left(\dot{D}_0(t) - \dot{D}_i(t)\right)}{\left(\max(\dot{D}_0(t))\right)}  \right)^2

%\begin{equation}
%m_2 = mean\left(	\frac{\left(\dot{D}_i^0 - \dot{D}_i^j\right)^2}{\left(\dot{D}_{max}^0\right)^2 } \right)
%\end{equation}
%
%\begin{equation}
%m_3 = max \left(	\frac{|\dot{D}_i^0 - \dot{D}_i^j|}{\dot{D}_{max}^0 } \right)
%\end{equation}
%
%\begin{equation}
%m_4 = max \left(	\frac{\left(\dot{D}_i^0 - \dot{D}_i^j\right)^2}{\left(\dot{D}_{max}^0\right)^2 } \right)
%\end{equation}
%
%\begin{equation}
%m_5 = mean\left( \frac{\nabla \dot{D}_i }{\dot{D}_i}\right)
%\end{equation}

To avoid the tails of the dose rate curves (where the dose rate is very low) affecting the computation of the metrics, the averaging was performed by excluding the tail beyond 5\% of the maximum dose rate of $\dot{D}_0(t)$.

For each radiopharmaceutical and for each curve $i$ we computed the TCP and the metrics accounting for the difference between curve $i$ and the reference curve. Then, we obtained the Spearman correlation coefficient between differences in TCP and the metric, and from this analysis, the metric that better correlated with differences in TCP was chosen.

Finally, the relationship between differences in TCP and the best metric (that with the highest association with changes in TCP) was investigated. These data were fitted and from these fits we obtained the value of the metric that leads to \emph{tolerable} differences in TCP (typically defined in this work as $\Delta$TCP=0.02), \emph{i.e.} the acceptable uncertainties in the determination of the dose rate curves.

\subsection[]{Implementation and parameters}

The models were implemented in MATLAB (Natick, MA, USA). The integrals to compute $G$ and the averaged TCP were numerically computed, avoiding negative values for any of the parameters involved. The computation of the double integral of Eq. (\ref{eq2}) can be time-consuming: in order to speed up the computation, we limited the computation of the second integral to,

\begin{equation}
\int_{t_{\rm min}}^t \frac{\dot{D}(t^\prime)}{D} e^{-\mu (t-t^\prime)} {\rm d}t^\prime	
\end{equation}
where $t_{\rm min}=\max(0, t-8\log(2)/\mu)$, is a value far enough from $t$ to verify that almost all the sublethal damage produced at $t_{\rm min}$ has been repaired at $t$. We checked numerically that this approximation did not introduce a bias in the calculation while importantly speeding up the process.

Regarding the generation of the dose rate curves discussed in section \ref{section_2_2}, different time discretizations ($\Delta t$)  were employed to solve the system (\ref{eq5}), depending on the radiopharmaceutical (the faster the decay the lower $\Delta t$) and the value of $\mu$ (the faster the repair the lower $\Delta t$): values ranged from $5\times10^{-3}$ to 0.1 h. Again, we checked numerically that the value of $\Delta t$ did not introduce a bias in the calculation.

The radiobiological parameters used for the calculations are presented in Table \ref{table_1}. We investigated the response of tumors with different $\alpha/\beta$ values (10 Gy, representative of most tumors, and 3 Gy, representative of tumors that are very sensitive to fractionation like prostate cancer). We also investigated different sub-lethal damage repair rates, ranging from 6.93 h$^{-1}$ (very fast repair with half-life of 6 minutes) to 0.23 h$^{-1}$ (slow repair with half-life of 3 h).

\begin{table}[htb]
    \captionv{14}{}{\label{table_1} List of parameter values used in this work.}
    \begin{center}
        \item[]\begin{tabular}{@{}ll}
            \hline
    	Parameter & Values\\
            \hline
           $\alpha$         & 0.2 Gy$^{-1}$ \\
           $\alpha/\beta$   & (3, 10) Gy \\
           $\mu$         &  (0.23, 0.35, 0.69, 1.39, 6.93) h$^{-1}$ \\
           $N$  & 10$^6$         \\
           \hline
        \end{tabular}
    \end{center}
\end{table}

\section[]{Results and Discussion}
\label{section_results}

\subsection[]{Dose rate curves}

\begin{figure}[H]
    \centering
    \includegraphics[width=12cm]{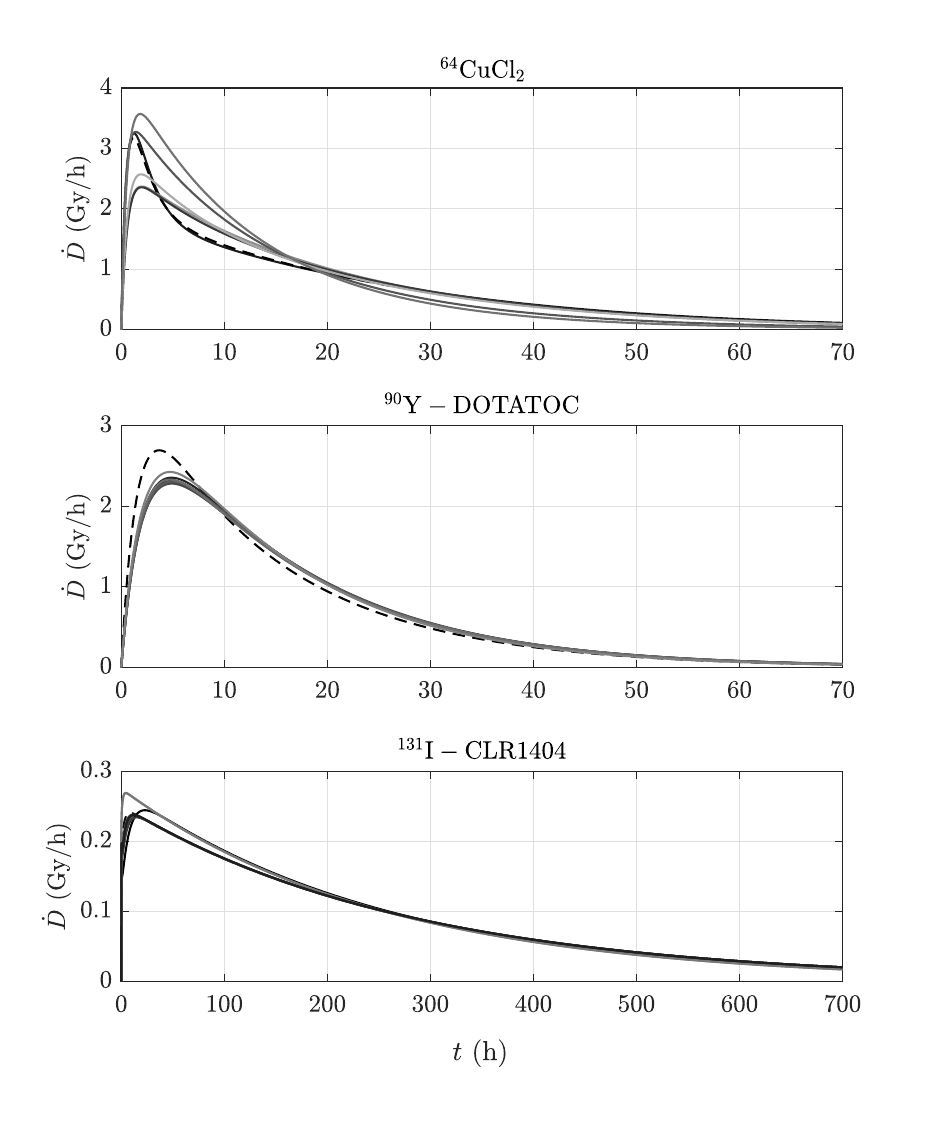}
    \captionv{14}{}{Dose rate curves for 131I-CLR1404 (10 curves), 90Y-DOTATOC (8 curves) and 64CuCl2 (7 curves) obtained by the methodology described in section \ref{section_2_2}. The curve acting as reference for each radiopharmaceutical is represented by a dashed line. For the curves represented in this figure, the normalization was performed by assuming a tumor with $\alpha/\beta$=10 Gy and $\mu$=0.69 h$^{-1}$ and imposing a TCP of 50\% for the reference curves.}
    \label{fig_2}
\end{figure}

We obtained 10, 8, and 7 curves for 131I-CLR1404, 90Y-DOTATOC and 64CuCl2, respectively, with the procedure to generate dose rate curves described in section \ref{section_2_2}. The shape of the dose rate curves is illustrated in Figure \ref{fig_2} for a tumor with $\alpha/\beta$=10 Gy and $\mu$=0.69 h$^{-1}$ (notice that for different combinations of tumor response parameters the dose rate curves were scaled up differently, as discussed in section \ref{section_2_2}, but their shape did not change). Overall, depending on the values of $\alpha/\beta$ and $\mu$ the total dose of the curves ranged from 54.7 to 71.3 Gy for 131I-CLR1404, 25.1 and 67.4 Gy for 90Y-DOTATOC, and 28.1 and 68.4 Gy for 64CuCl2.

\subsection[]{Correlation between TCP and dose rate differences}

Metrics 1-3 and TCP values were computed for each radiopharmaceutical and each curve. A Spearman correlation test between metric values and differences in TCP ($| \Delta\mathrm{TCP}_i | = | \mathit{TCP}_{\mathrm{ref}}-\mathit{TCP}_i|$)  was performed to investigate correlations. The average correlation coefficients obtained for each metric (average of the results obtained for each radiopharmaceutical and each combination of radiobiological parameters) showed that metric 1 clearly correlated the best with the differences in TCP, with an average correlation coefficient of 0.720 (versus 0.455 for metric 2 and 0.311 for metric 3), and was therefore selected for the rest of this study.

%\begin{table}[tbh]
%\renewcommand{\arraystretch}{1.5} 
%\centering
%\begin{tabular}{|c|c|c|c|c|c|}
%\hline
%Metric            & $m_1$ & $m_2$ & $m_3$  \\ \hline
%$\overline{\rho}$ & 0.720        & 0.455       & 0.311     \\ \hline
%\end{tabular}
%\caption{Average Spearman's rank correlation coefficients (average of the results obtained for each radiophamarceutical and each combination of radiobiological parameters) between dose rate variations quantified with different metrics and variations of TCP.}
%\label{table2}
%\end{table}

\begin{figure}[tb]
\centering
%\hspace*{-3cm} 
\includegraphics[width=15cm]{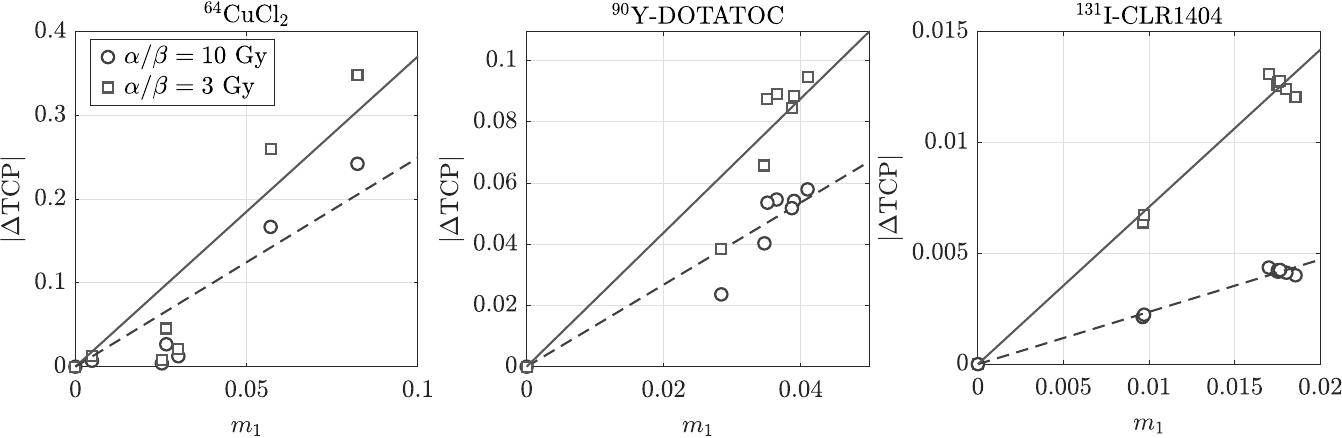}
\captionv{14}{}{Scatter plot of $| \Delta\mathrm{TCP} |$ vs. $m_1$ for $\mu$ = 0.69 h$^{-1}$, and $\alpha/\beta$=3 and 10 Gy. Linear fits of the data for $\alpha/\beta$=3 Gy (dashed lines) and $\alpha/\beta$=10 Gy (dash-dotted lines) are also shown.}
\label{fig_3}
\end{figure}

In Figure \ref{fig_3} we illustrate the association between variations in TCP and the value of the metric $m_1$ accounting for differences in the dose rate curves by presenting scatter plots for the three radiopharmaceuticals under study,  with $\alpha/\beta$=10 and 3 Gy and $\mu$=0.69 h$^{-1}$.

After selecting the optimal metric, fits of $| \Delta\mathrm{TCP} |$ against the value of this metric were performed for each of the parameter combinations, as shown in Figure \ref{fig_3}: linear fits of the form $|\Delta\mathrm{TCP} | = b  m_1$ (no independent term) were chosen because the scatter plots qualitatively showed a linear relationship between both magnitudes, and $|\Delta\mathrm{TCP}| = 0$ is expected for $m_1=0$. From such fits, we computed the value of the metric corresponding to a maximum $| \Delta\mathrm{TCP} |$ of our choice, which was set to 0.02, simply as $m_{1,\: 0.02}=0.02/b$. This value quantifies the sensitivity of the TCP to changes in the dose-rate profile: a lower value $m_{1,\: 0.02}$ means that small changes in the dose rate profile lead to $2\%$ variations on the TCP, while the opposite holds for higher values. Error propagation analysis was used on the computation of $m_{1,\: 0.02}$ to obtain the uncertainties associated to it arising from the fit uncertainties.

\begin{figure}[tb]
\centering
%\hspace*{-3cm} 
\includegraphics[width=15cm]{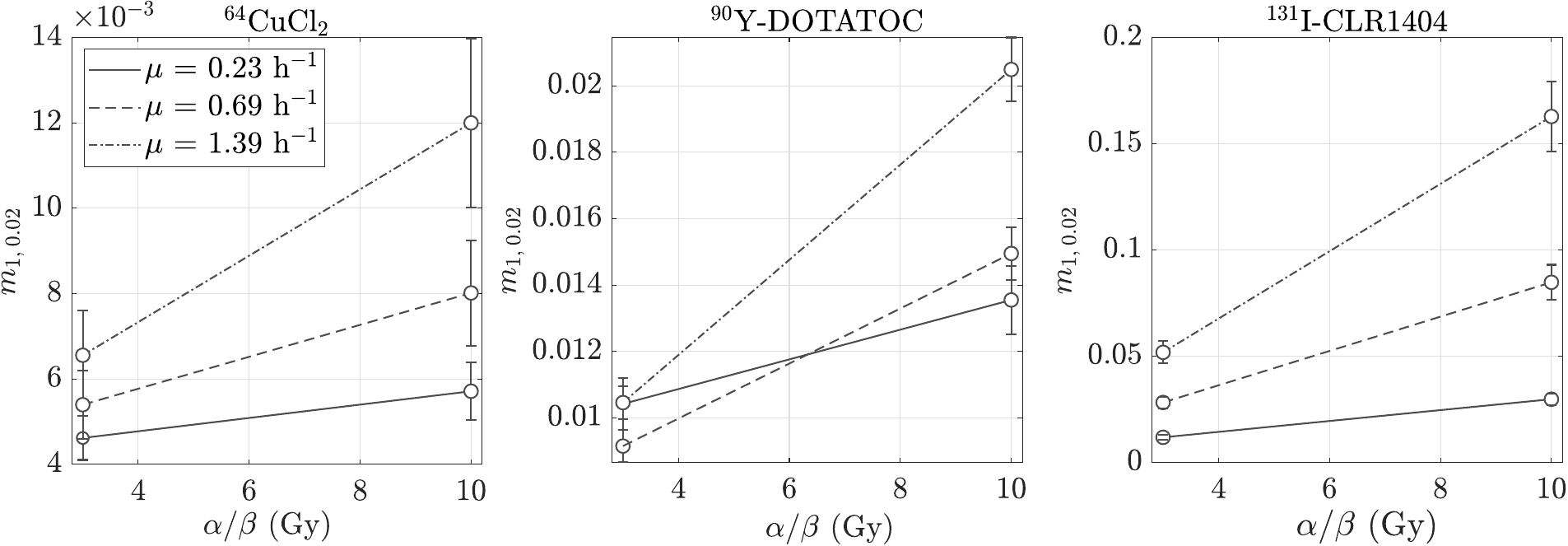}
\captionv{14}{}{Dependence of the sensitivity of the TCP to changes in the dose rate curve, quantified through the parameter $m_{1,\: 0.02}$, on the $\alpha/\beta$ of tumor cells. We report results for each radiopharmaceutical and for different sub-lethal damage repair rates.}
\label{fig_4}
\end{figure}

\begin{figure}[tb]
\centering
%\hspace*{-3cm} 
\includegraphics[width=15cm]{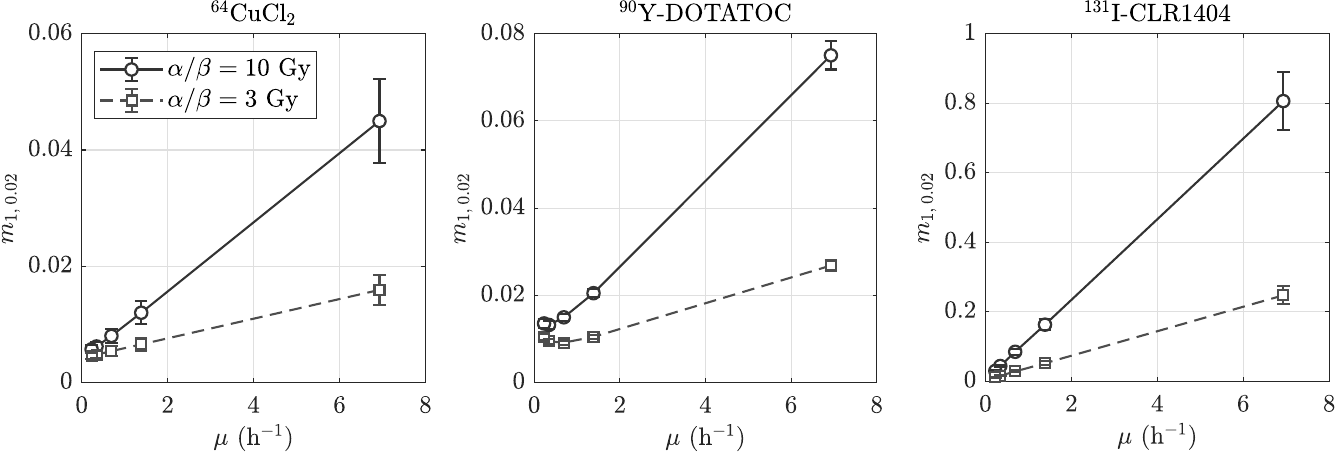}
\captionv{14}{}{Dependence of the sensitivity of the TCP to changes in the dose rate curve, quantified through the parameter $m_{1,\: 0.02}$, on the sub-lethal damage repair rate of tumor cells ($\mu$). We report results for each radiopharmaceutical and for different $\alpha/\beta$ values.}
\label{fig_5}
\end{figure}

\begin{figure}[tb]
\centering
\includegraphics[width=10cm]{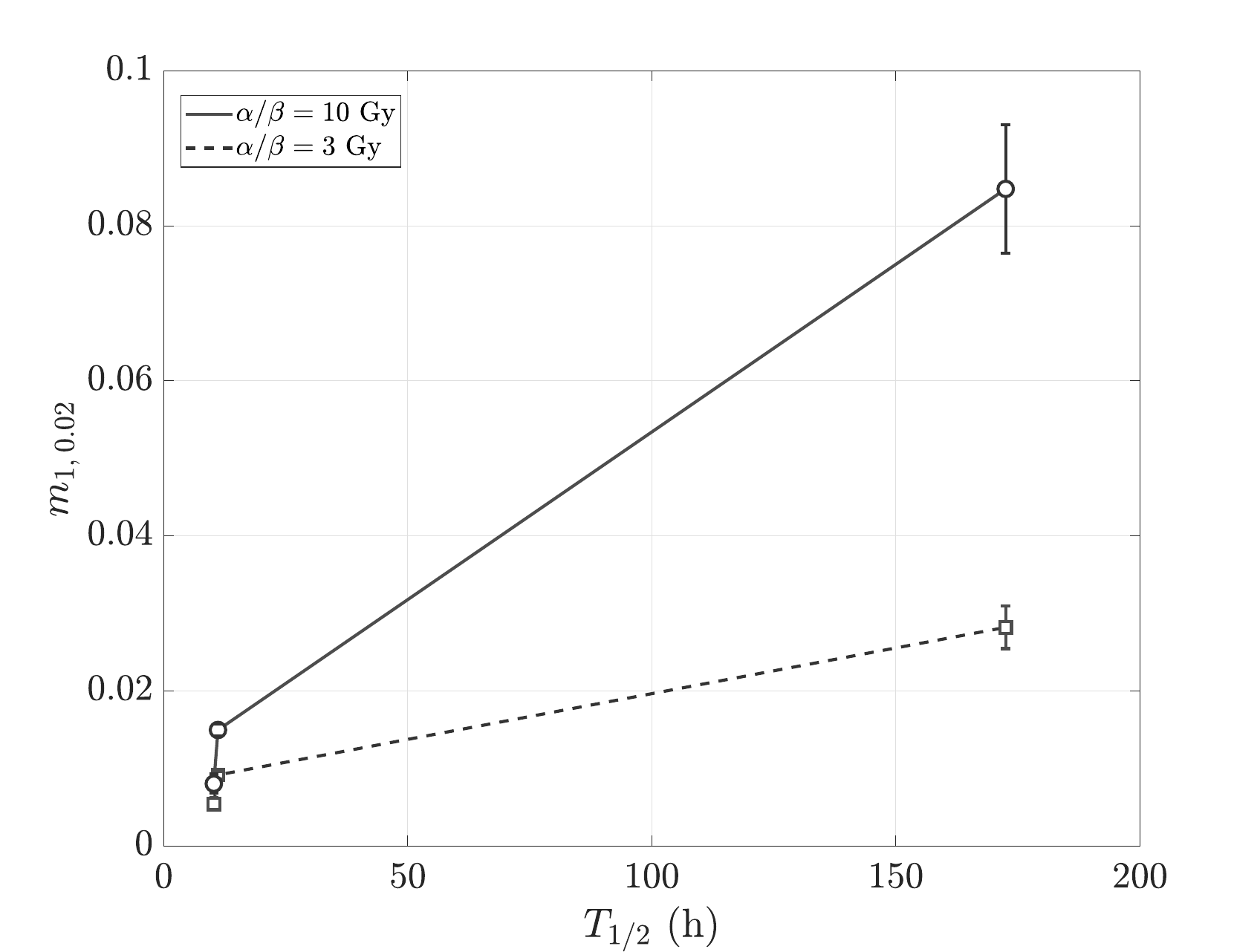}
\captionv{14}{}{Dependence of the sensitivity of the TCP to changes in the dose rate curve, quantified through the parameter $m_{1,\: 0.02}$, on the biological half-life of the radiopharmaceuticals. We report results for $\alpha/\beta$=3 and 10 Gy, and $\mu$=0.69  h$^{-1}$.}
\label{fig_6}
\end{figure}

In Figures \ref{fig_4}, \ref{fig_5} and \ref{fig_6} we illustrate the dependence of $m_{1,\: 0.02}$ with the $\alpha/\beta$ and sub-lethal repair rate of tumor cells, and with the biological half-life of each radiopharmaceutical (average of the half-life of each curve, calculated as the time taken to decay from maximum to half-maximum dose rate). 

These results show that the sensitivity of the TCP to dose rate variations decreases (\emph{i.e.} larger values of $m_{1,\: 0.02}$) with increasing $\alpha/\beta$ and sub-lethal damage repair rate of the tumor cells, and increasing biological half-life of the dose rate curves. The former effect is expected, as according to classical radiobiology the dose rate effect is included in the quadratic term of the LQ model. A larger $\alpha/\beta$ ratio implies a lower $\beta$, a lower contribution of the quadratic term, and therefore a lower dose rate effect.

The interpretation of the other dependences is somewhat more complex. The dose rate effect is expected to disappear both in the limits $G\to 0$ and $G\to 1$. The decrease of the sensitivity of the TCP to dose rate variations observed in Figures \ref{fig_5} and \ref{fig_6} for increasing sub-lethal damage repair rate and biological half-life of the dose rate curve corresponds to the former limit ($G\to 0$ when $\mu\to \infty$ and $T_{1/2}\to \infty$). However, in the limit of instantaneous dose delivery (which can be achieved with either $\mu\to 0$ or $T_{1/2}\to 0$), $G\to 1$, the dose effect would disappear, and therefore the sensitivity to dose rate variations would also disappear resulting in increasing values of $m_{1,\: 0.02}$ with decreasing values of sub-lethal damage repair rate and biological half-life of the dose rate curves. This would lead to inverted bell curves which are not observed for the values of $\mu$ and $T_{1/2}$ investigated in this work.

\begin{figure}[tb]
\centering
\includegraphics[width=12cm]{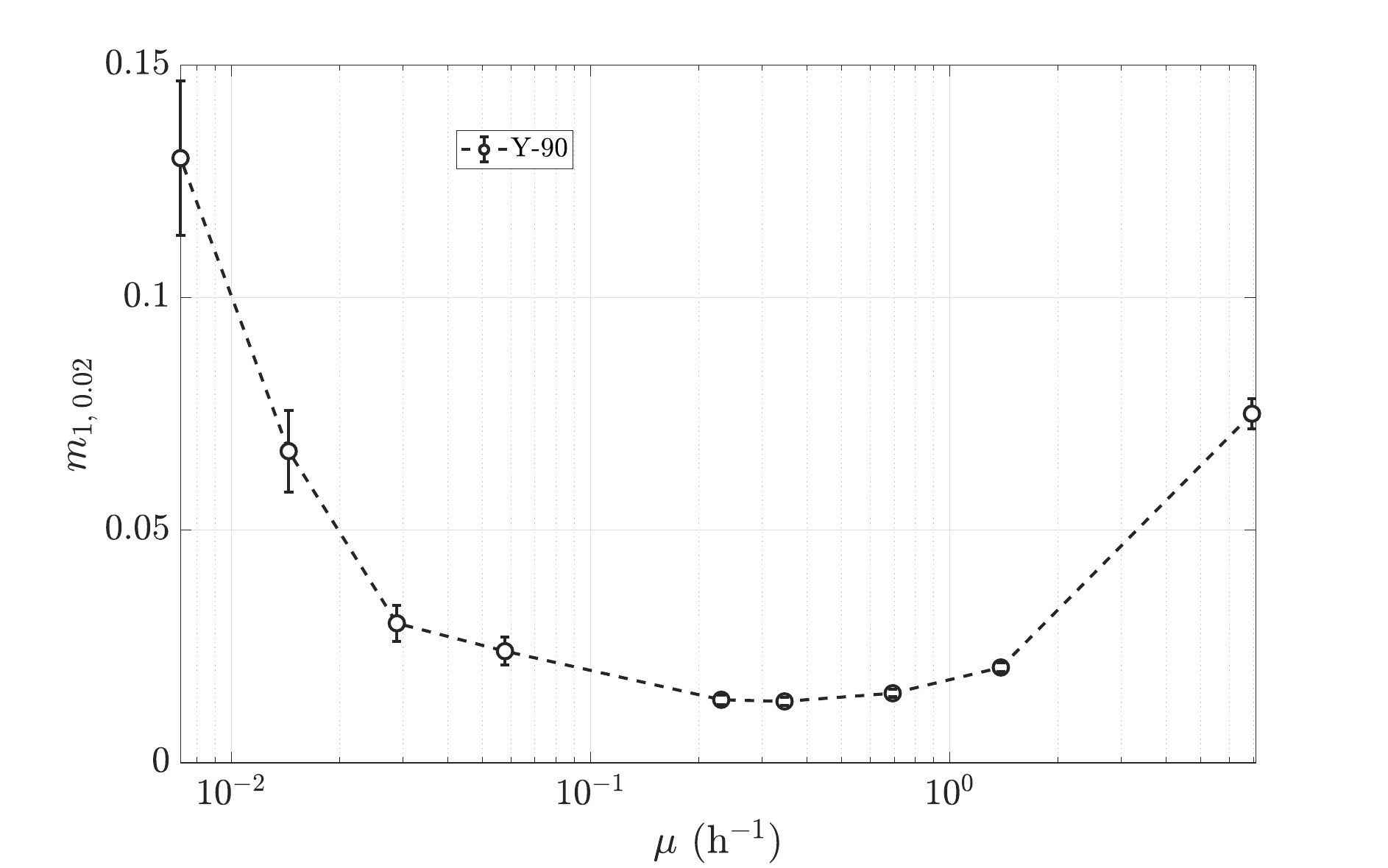}
\captionv{14}{}{Dependence of the sensitivity of the TCP to changes in the dose rate curve, quantified through the parameter $m_{1,\: 0.02}$, on the sub-lethal damage repair rate of tumor cells ($\mu$), for 90Y-DOTATOC and $\alpha/\beta$=10 Gy. Values of $\mu$ well below those investigated for Figure \ref{fig_5} were investigated to explore the limit $G\to 1$. Notice the log scale in the x-axis to better appreciate the shape of the curve in the limits $\mu\to 0$ ($G\to 1$) and $\mu\to \infty$ ($G\to 0$) }
\label{fig_7}
\end{figure}

In order to qualitatively explore this behaviour in the limit $G\to 1$, we extended the analysis of the dependence of the sensitivity of the TCP on sub-lethal damage repair rate by including faster repair rates, down to $\sim$0.01 h$^{-1}$. While such fast repairs are certainly not biologically achievable, this study allowed to fully investigate the shape of the TCP sensitivity curve in the limit $G\to 1$. This extended study was performed only for 90Y-DOTATOC, and the results are reported in Figure \ref{fig_7}.

\subsection{Limitations of this work}

This piece of theoretical work has several limitations that may affect the results here obtained. Therefore, the results here obtained linking the sensitivity of the TCP to dose rate uncertainties and the dependence of this sensitivity on radiopharmaceutical kinetics and radiobiological parameters should be considered qualitative rather than quantitative.

Among the limitations, the repair of sub-lethally damaged cells was considered as exponential, but other functional forms like e.g. Michaelis–Menten or bi-exponential kinetics may be more realistic \cite{murray2003,dale2019}. Nonetheless, more complex repair kinetics could be easily included in the framework presented in this work.

We have also investigated single-cycle MRT treatments and ignored tumor proliferation during treatment. Including multi-cycle MRT treatment would require modelling tumor proliferation in between cycles, for example by using more complex differential models like that used in \cite{galler2024} to simulate multiple cycles of $^{177}$Lu MRT.

Another limitation is the assumption that the absorbed dose in the tumor is homogeneous for the computation of tumor control probabilities. In clinical treatments, absorbed dose in the tumor may present important macroscopic heterogeneities, which could be included in the LQ-Poisson model by using differential dose volume histograms \cite{webb1993}. Also, heterogeneities at the subcelular level can affect tumor control\cite{uusijarvi2008}. We have opted against including such level of complication for the sake of simplicity and because of the lack of reliable clinical studies on dose heterogeneities. Also, while according to classical radiobiology, the tumor control probability would strongly depend on the minimum dose in the tumor for highly heterogeneous dose distributions, this paradigm has been challenged by the experimental on the abscopal effect and response to spatially fractionated radiotherapy. Modelling tumor control probability for heterogeneous dose distributions may require novel models beyond classical radiobiology  \cite{mcmillan2024, potiron2024}.

\section[Conclusions]{Conclusions}
\label{section_conclusions}

In this work, we have investigated how sensitive the effectiveness (tumor control probability) of a MRT treatment is to the dose rate effect. Based on this study, we have obtained metrics that correlate the difference between two dose rate curves and their effectiveness. Such metrics can be used to estimate the maximum uncertainties allowed in the reconstruction of a dose rate curve in order to limit the uncertainty of the effectiveness below a given value, a procedure that has been illustrated by limiting differences in TCP to $\le$ 0.02.

We have found that the type of radiopharmaceutical (due to different decay/clearance rates) and the tumor response parameters different ($\alpha/\beta$, sub-lethal repair rate) can importantly affect the dose-rate effect on the treatment effectiveness and therefore the sensitivity to dose rate uncertainties.

The radiobiological effect of a MRT treatment on a tumor depends not only on the absorbed dose but also on the dose rate profile. Ideally, both magnitudes should be measured with accuracy in order to progress towards the optimization of individualized MRT treatments based on the \emph{dose (and dose rate)-effect} relationship, similarly to external radiotherapy. Our study suggests that this would be more important for tumors with low $\alpha/\beta$ and moderately slow sub-lethal damage repair ($\sim$0.5 h$^{-1}$) treated with fast-decaying radiopharmaceuticals.

\section*{Acknowledgements}

This work has received funding from Xunta de Galicia-GAIN (IN607D 2022/02) and Ministerio de Ciencia e Innovación, Agencia Estatal de Investigación and FEDER, UE (grants PID2021-128984OB-I00 and PLEC2022-009476).

\section*{Conflicts of interest}

The authors have no relevant conflicts of interest to disclose.

\addcontentsline{toc}{section}{\numberline{}References}

% The Following assumes you are using Bibtex. However, for submission to the
% journal you MUST explicitly INCLUDE THE REFERENCES IN THE TEX FILE. 
% In that case you need the following

% \begin{thebibliography}{10}
% Insert the .bbl file generated by BibTeX here
	%This will be a series of entries from your .bib file formatted
	%something like
	%\bibitem{Me09}
        %{I.~Meijsing, B.~W.~Raaymakers, A.~J.~E.~Raaijmakers \it et~al.},
        %\newblock {Dosimetry for the MRI accelerator: the impact of a 
	%magnetic field on the response of a Farmer NE2571 ionization chamber},
        %\newblock Phys. Med. Biol. {\bf 54}, 2993 -- 3002 (2009).
% \end{thebibliography}

% The following is when using BibTeX and picks up the example.bib file
%\bibliography{refs.bib}

\begin{thebibliography}{10}

\bibitem{sgouros2020}
G.~Sgouros, L.~Bodei, M.~R. McDevitt, and J.~R. Nedrow,
\newblock Radiopharmaceutical therapy in cancer: clinical advances and
  challenges,
\newblock Nature Reviews Drug Riscovery {\bf 19}, 589--608 (2020).

\bibitem{silberstein2012}
E.~B. Silberstein et~al.,
\newblock The SNMMI practice guideline for therapy of thyroid disease with 131I
  3.0,
\newblock Journal of Nuclear Medicine {\bf 53}, 1633--1651 (2012).

\bibitem{fendler2017}
W.~P. Fendler, K.~Rahbar, K.~Herrmann, C.~Kratochwil, and M.~Eiber,
\newblock 177Lu-PSMA radioligand therapy for prostate cancer,
\newblock Journal of Nuclear Medicine {\bf 58}, 1196--1200 (2017).

\bibitem{cwikla2010}
J.~Cwikla, A.~Sankowski, N.~Seklecka, J.~Buscombe, A.~Nasierowska-Guttmejer,
  K.~Jeziorski, R.~Mikolajczak, D.~Pawlak, K.~Stepien, and J.~Walecki,
\newblock Efficacy of radionuclide treatment DOTATATE Y-90 in patients with
  progressive metastatic gastroenteropancreatic neuroendocrine carcinomas
  (GEP-NETs): a phase II study,
\newblock Annals of Oncology {\bf 21}, 787--794 (2010).

\bibitem{davis2023}
L.~Davis, C.~Elmaraghi, J.~R. Buscombe, and M.~N. Gaze,
\newblock Clinical perspectives on dosimetry in molecular radiotherapy,
\newblock Physica Medica {\bf 114}, 103154 (2023).

\bibitem{stokke2017}
C.~Stokke et~al.,
\newblock Dosimetry-based treatment planning for molecular radiotherapy: a
  summary of the 2017 report from the Internal Dosimetry Task Force,
\newblock EJNMMI Physics {\bf 4}, 1--9 (2017).

\bibitem{ivashchenko2024}
O.~V. Ivashchenko, J.~O'Doherty, D.~Hardiansyah, M.~Cremonesi, J.~Tran-Gia,
  E.~Hippel{\"a}inen, C.~Stokke, E.~Grassi, M.~Sandstr{\"o}m, and G.~Glatting,
\newblock Time-activity data fitting in molecular radiotherapy: methodology and
  pitfalls,
\newblock Physica Medica {\bf 117}, 103192 (2024).

\bibitem{strigari2014}
L.~Strigari, M.~Konijnenberg, C.~Chiesa, M.~Bardies, Y.~Du, K.~S. Gleisner,
  M.~Lassmann, and G.~Flux,
\newblock The evidence base for the use of internal dosimetry in the clinical
  practice of molecular radiotherapy,
\newblock European Journal of Nuclear Medicine and Molecular Imaging {\bf 41},
  1976--1988 (2014).

\bibitem{jeremic2018}
M.~Z. Jeremic, M.~D. Matovic, D.~Z. Krstic, S.~B. Pantovic, and D.~R. Nikezic,
\newblock A five-compartment biokinetic model for 90Y-DOTATOC therapy,
\newblock Medical Physics {\bf 45}, 5577--5585 (2018).

\bibitem{guiu2018}
J.~Guiu-Souto, S.~Neira-Castro, M.~S{\'a}nchez-Garc{\'\i}a, O.~L. Pouso,
  M.~Pombar-Came{\'a}n, and J.~Pardo-Montero,
\newblock Adaptive biokinetic modelling of iodine-131 in thyroid cancer
  treatments: implications on individualised internal dosimetry,
\newblock Journal of Radiological Protection {\bf 38}, 1501 (2018).

\bibitem{neira2021}
S.~Neira, A.~Gago-Arias, I.~G{\'o}nzalez-Crespo, J.~Guiu-Souto, and
  J.~Pardo-Montero,
\newblock Development of a Compartmental Pharmacokinetic Model for Molecular
  Radiotherapy with 131I-CLR1404,
\newblock Pharmaceutics {\bf 13}, 1497 (2021).

\bibitem{kost2015}
S.~D. Kost, Y.~K. Dewaraja, R.~G. Abramson, and M.~G. Stabin,
\newblock VIDA: a voxel-based dosimetry method for targeted radionuclide
  therapy using Geant4,
\newblock Cancer Biotherapy \& Radiopharmaceuticals {\bf 30}, 16--26 (2015).

\bibitem{marcatili2013}
S.~Marcatili, C.~Pettinato, S.~Daniels, G.~Lewis, P.~Edwards, S.~Fanti, and
  E.~Spezi,
\newblock Development and validation of RAYDOSE: a Geant4-based application for
  molecular radiotherapy,
\newblock Physics in Medicine \& Biology {\bf 58}, 2491--2508 (2013).

\bibitem{besemer2018}
A.~E. Besemer, Y.~M. Yang, J.~J. Grudzinski, L.~T. Hall, and B.~P. Bednarz,
\newblock Development and validation of RAPID: a patient-specific Monte Carlo
  three-dimensional internal dosimetry platform,
\newblock Cancer Biotherapy \& Radiopharmaceuticals {\bf 33}, 155--165 (2018).

\bibitem{neira2020b}
S.~Neira et~al.,
\newblock Quantification of internal dosimetry in PET patients: individualized
  Monte Carlo vs generic phantom-based calculations,
\newblock Medical Physics {\bf 47}, 4574--4588 (2020).

\bibitem{morganti2021}
S.~Morganti, F.~Collamati, R.~Faccini, G.~Iaccarino, C.~Mancini-Terracciano,
  R.~Mirabelli, F.~Nicolanti, M.~Pacilio, A.~Soriani, and
  E.~Solfaroli-Camillocci,
\newblock A wearable radiation measurement system for collection of
  patient-specific time-activity data in radiopharmaceutical therapy: system
  design and Monte Carlo simulation results,
\newblock Medical Physics {\bf 48}, 8117--8126 (2021).

\bibitem{mancini2023}
C.~Mancini-Terracciano et~al.,
\newblock Experimental validation of an innovative approach in biokinetics
  study for personalised dosimetry of molecular radiation therapy treatments,
\newblock Physics in Medicine \& Biology {\bf 68}, 19NT02 (2023).

\bibitem{dale1996}
R.~Dale,
\newblock Dose-rate effects in targeted radiotherapy,
\newblock Physics in Medicine \& Biology {\bf 41}, 1871--1884 (1996).

\bibitem{dale2005}
R.~Dale and A.~Carabe-Fernández,
\newblock The radiobiology of conventional radiotherapy and its application to
  radionuclide therapy,
\newblock Cancer Biotherapy \& Radiopharmaceuticals {\bf 20}, 47--51 (2005).

\bibitem{orourke2009}
S.~O’Rourke, H.~McAneney, and T.~Hillen,
\newblock Linear quadratic and tumour control probability modelling in external
  beam radiotherapy,
\newblock Journal of Mathematical Biology {\bf 58}, 799--817 (2009).

\bibitem{steel1986}
G.~G. Steel, J.~D. Down, J.~H. Peacock, and T.~C. Stephens,
\newblock Dose-rate effects and the repair of radiation damage,
\newblock Radiotherapy and Oncology {\bf 5}, 321--331 (1986).

\bibitem{dale1985}
R.~G. Dale,
\newblock The application of the linear-quadratic dose-effect equation to
  fractionated and protracted radiotherapy,
\newblock The British Journal of Radiology {\bf 58}, 515--528 (1985).

\bibitem{hall1991}
E.~J. Hall and D.~J. Brenner,
\newblock The dose-rate effect revisited: radiobiological considerations of
  importance in radiotherapy,
\newblock International Journal of Radiation Oncology* Biology* Physics {\bf
  21}, 1403--1414 (1991).

\bibitem{dale1999}
R.~G. Dale,
\newblock A new incomplete-repair model based on a 'reciprocal-time' pattern of
  sublethal damage repair,
\newblock Acta Oncologica {\bf 38}, 919--929 (1999).

\bibitem{galler2024}
M.~Galler, C.~Chibolela, F.~Thiele, J.~M.~M. Rogasch, and H.~Amthauer,
\newblock Dose-rate effects and tumor control probability in 177{Lu}-based
  targeted radionuclide therapy: a theoretical analysis,
\newblock Physics in Medicine \& Biology {\bf 69}, 205003 (2024).

\bibitem{webb1993}
S.~Webb and A.~E. Nahum,
\newblock A model for calculating tumour control probability in radiotherapy
  including the effects of inhomogeneous distributions of dose and clonogenic
  cell density,
\newblock Physics in Medicine \& Biology {\bf 38}, 653--666 (1993).

\bibitem{besemer2019}
A.~E. Besemer, J.~J. Grudzinski, J.~P. Weichert, L.~T. Hall, and B.~P. Bednarz,
\newblock Pretreatment CLR 124 positron emission tomography accurately predicts
  CLR 131 three-dimensional dosimetry in a triple-negative breast cancer
  patient,
\newblock Cancer Biotherapy \& Radiopharmaceuticals {\bf 34}, 13--23 (2019).

\bibitem{righi2018}
S.~Righi et~al.,
\newblock Biokinetic and dosimetric aspects of 64CuCl2 in human prostate
  cancer: possible theranostic implications,
\newblock EJNMMI Research {\bf 8}, 1--9 (2018).

\bibitem{murray2003}
J.~Murray,
\newblock {\em Mathematical Biology II: Spatial Models and Biomedical
  Applications},
\newblock Springer New York, 2003.

\bibitem{dale2019}
R.~G. Dale,
\newblock Radiation repair models for clinical application,
\newblock The British Journal of Radiology {\bf 92}, 20180070 (2019).

\bibitem{uusijarvi2008}
H.~Uusij{\"a}rvi, P.~Bernhardt, and E.~Forssell-Aronsson,
\newblock Tumour control probability (TCP) for non-uniform activity
  distribution in radionuclide therapy,
\newblock Physics in Medicine \& Biology {\bf 53}, 4369--4381 (2008).

\bibitem{mcmillan2024}
M.~T. McMillan, A.~J. Khan, S.~N. Powell, J.~Humm, J.~O. Deasy, and
  A.~Haimovitz-Friedman,
\newblock Spatially Fractionated Radiotherapy in the Era of Immunotherapy,
\newblock Seminars in Radiation Oncology {\bf 34}, 276--283 (2024).

\bibitem{potiron2024}
S.~Potiron et~al.,
\newblock The significance of dose heterogeneity on the anti-tumor response of
  minibeam radiation therapy,
\newblock Radiotherapy and Oncology {\bf 201}, 110577 (2024).

\end{thebibliography}

% following defines the style of the .bbl file 
%\bibliographystyle{medphy.bst}

% Note that you need to typeset once, then run Bibtex, then typeset another
% two times to get the references working properly.

\end{document}